\documentclass[runningheads,a4paper]{llncs}

\usepackage{amssymb}
\usepackage{graphicx}

\usepackage{wrapfig}
\usepackage{url}
\usepackage{booktabs}
\usepackage{array}
\usepackage{times}   
\usepackage{multirow}

\usepackage{microtype}
\usepackage{url}

\newcommand{\keywords}[1]{\par\addvspace\baselineskip
\noindent\keywordname\enspace\ignorespaces#1}

\newcommand\Mark[1]{\textsuperscript#1}

\begin{document}

\mainmatter  

\title{Storytelling Agents with Personality and Adaptivity}

\titlerunning{Storytelling Agents with Personality and Adaptivity}

\toctitle{Storytelling Agents with Personality and Adaptivity}
\author{\Mark{1}Zhichao Hu, \Mark{1}Marilyn A. Walker, \Mark{2}Michael Neff and \Mark{1}Jean E. Fox Tree}
\authorrunning{Zhichao Hu, Marilyn A. Walker, Michael Neff and Jean E. Fox Tree} 
\institute{\Mark{1}University of California, Santa Cruz~~~~~~~~~~~~~~~~~~\Mark{2}University of California, Davis\\
\email{\{zhu,mawalker,foxtree\}@ucsc.edu~~~~~mpneff@ucdavis.edu}}

\maketitle

\vspace{-0.2in}
\begin{abstract}
We explore the expression of personality and adaptivity
through the gestures of virtual agents in a storytelling
task. We conduct two experiments using four different dialogic stories.
We manipulate agent personality on the extraversion scale,
whether the agents adapt to one another in their
gestural performance and agent gender.
Our results show that subjects are
able to perceive the intended variation in extraversion
between different virtual agents, independently of the story they
are telling and the gender of the agent. A second study shows that 
subjects also prefer
adaptive to nonadaptive virtual agents.
\end{abstract}

\section{Introduction}
\label{intro-sec}

It is a truism that every person is a unique individual. However, when
interacting with or observing others, people make inferences that
generalize from specific, observed behaviors to explanations for those
behaviors in terms of dispositional traits
\cite{nisbett1980trait}.
One theory that attempts to account for such inferences is the Big
Five theory of personality, which posits that consistent
patterns in the way individuals behave, feel, and think across
different situations, can be described in terms of trait adjectives,
such as sociable, shy, trustworthy, disorganized or imaginative
\cite{Mehletal06,Norman63}. 

Previous work suggests both that personality traits are {\em real},
and that they are {\em useful} as a basis for models for Intelligent
Virtual Agents (IVAs) for a range of applications
\cite{BickmoreSchulman06,Hartmann2005,kopp04,thie08,helo09}. Many
findings about how people perceive other humans appear to carry over
to their perceptions of IVAs
\cite{Andreetal99,Ruttkayetal04,MoonNass96,DBLP:journals/aamas/EndrassARN13,vhp-noma2000design}.
Research suggests that human users are more engaged and thus learn
more when interacting with characters endowed with personality and
emotions, and that a character's personality, surprisingly, affects
users' perceptions of the system's competence
\cite{tapus2008a,Wangetal05}.  Recent experiments show
that the Big Five theory is a useful basis for multi-modal
integration of nonverbal and linguistic behavior, and that
automatically generated variations in personality are perceived as
intended \cite{MairesseWalker11,Neffetal10,Beeetal10,Neffetal11}.

However, personality is not expressed in a void.  
Conversants dynamically adapt to their conversational
partner, both in conversation and when telling stories, and using both
verbal and nonverbal features
\cite{Foxtree99,ParrillKimbara06,BailensonNick05,tolins2014addressee,thorne2007channeling},
     {\it inter alia}.  There is also evidence that people prefer IVAs
     that align with human behavior, such as by mimicking head
     movements \cite{BailensonNick05} or speech style
     \cite{MoonNass96}. A human's attraction to
     an IVA is increased when the IVA adapts its personality to the
     human over time rather than maintaining a consistently similar
     personality \cite{MoonNass96}. Inspired by previous work, this paper:

\begin{itemize}
\item Introduces a novel task of two IVAs co-telling a story;
\item Varies IVA personality through gestural parameters of gesture rate, speed,
expanse and form.
\item Varies whether the IVAs adapt to one another's gestures in gesture rate, speed, expanse and form and use of specific gestures.
\item Tests the effect of, and interaction between, these
variations with human perceptual experiments and report our results.
\end{itemize}

\begin{wrapfigure}{r}{2.5in}
\vspace{-0.7in}
\begin{center}
\begin{scriptsize}  
\begin{tabular}{|rp{2.3in}|}
\hline
\multicolumn{2}{|c|}{\bf Protest Story}  \\
\hline
A1: &  Hey, do you remember that day? It was a work day, I remember there was some big event going on. \\

B1: &  Yeah, that day was the start of the G20 summit. It's an event that happens every year. \\

A2: &  Oh yeah, right, it's that meeting where 20 of the leaders of the world come together. They talk about how to run their governments effectively. \\

B2: &  Yeah, exactly. There were many leaders coming together. They had some pretty different ideas about what's the best way to run a government. \\ 

A3: &  And the people who follow the governments also have different ideas. Whenever world leaders meet, there will be protesters expressing different opinions. I remember the protest that happened just along the street where we work. \\ 

B3: &  It looked peaceful at the beginning.... \\

A4: &  Right, until a bunch of people started rebelling and creating a riot. \\

B4: &  Oh my gosh, it was such a riot, police cars were burned, and things were thrown at cops. \\

A5: &  Police were in full riot gear to stop the violence. \\

B5: &  Yeah, they were. When things got worse, the protesters smashed the windows of stores. \\

A6: &  Uh huh. And then police fired tear gas and bean bag bullets. \\

B6: &  That's right, tear gas and bean bag bullets... It all happened right in front of our store. \\

A7: &  That's so scary. \\

B7: &  It was kind of scary, but I had never seen a riot before, so it was kind of interesting for me.\\
\hline 
\end{tabular}
\caption{Protest Dialogue, with fixed level of linguistic
  adaptation. \label{protest_blog_dialog}}
\end{scriptsize}  
\end{center}
\vspace{-0.4in}
\end{wrapfigure}

Our stories come from weblogs of personal narratives~\cite{GordonSwanson09} 
whose content has been regenerated as dialogues to support story
co-telling.  Example dialogs from the four we use in our
experiments are in Fig.~\ref{protest_blog_dialog}
and Fig.~\ref{pet_blog_dialog} 
in Sec.~\ref{exp-sec}.  These
dialogs have a fixed linguistic representation and use oral language,
discourse markers, shorter sentences, and repetitions and
confirmations between speakers, as well as techniques to make the
story sound like the two speakers experienced the event together.  Our
aim is to mimic the finding that storytelling in the wild is naturally
conversational \cite{thorne2007channeling},
and that the style of oral storytelling 
among friends varies depending 
on their personalities \cite{thorne2007channeling}. 

We carry out two experiments. In the personality experiment, we elicit
subjects' perceptions of two virtual agents designed to have
different personalities. In the gestural adaptation experiment, we ask
whether subjects prefer adaptive vs. non adaptive
agents.  Our results show that agents intended to be extroverted
or introverted 
are perceived as such, and that subjects prefer adaptive stories.
Sec.~\ref{corpus-sec} describes our story dialog corpus.
Sec.~\ref{exp-sec} and \ref{res-sec} presents our experimental 
design and results.  In order to compare more concisely with our 
work, we delay discussion of related work until
Sec.~\ref{conc-sec}, where we discuss our results and describe future
work.

\section{Story Dialog Corpus}
\label{corpus-sec}

We first annotate dialogs with 
a general underspecified gesture representation, then we prepare
several versions of each dialog by varying experimental parameters
such as agent extraversion and adaptivity.

\subsubsection{Gesture Annotation}
We build on Neff et al.'s work on the impact of extraversion on gesture in
IVAs~\cite{Neffetal10}, as shown in
Table~\ref{tab-extraversion-gesture}, and select parameters
to depict both introverted
and extraverted IVAs by varying gesture
amplitude, direction, rate and speed.  We test user perceptions
of IVA personality during story co-telling without adaptation, and
then test whether we can achieve effects on personality perception
when IVAs adapt to one another.

\begin{table}[htb]  
\centering  
\begin{scriptsize}  
\begin{tabular}{|>{\raggedright}p{1.2in}|p{1in}|p{2.3in}|}  
\hline  
{\bf Parameter} & {\bf Introvert Findings} & {\bf Extravert Findings} \\  
\hline 
\hline \textbf{Gesture amplitude} &     narrow  &       wide, broad \\  
  
\hline \textbf{Gesture direction} &     inward, self-contact    &       outward, table-plane and horizontal spreading gesture \\  
\hline \textbf{Gesture rate} &  low     & high, more movements of head, hands and legs\\ 
  
\hline \textbf{Gesture speed, response time} & slow &   fast, quick     \\  
  
\hline \textbf{Gesture connection} & low smoothness, rhythm disturbance &       smooth, fluent \\  
\hline \textbf{Body part} &   &  head tilt, shoulder erect, chest forward, limbs spread, elbows away from body, hands away from body, legs apart, legs leaning, bouncing, shaking of legs             \\  
\hline  
\end{tabular}  
\end{scriptsize}  
\caption{The gestural correlates of extraversion. \label{tab-extraversion-gesture} }
\vspace{-0.25in}
\end{table}

\begin{figure}[t!b]
\begin{center}
\includegraphics[width=4.75in]{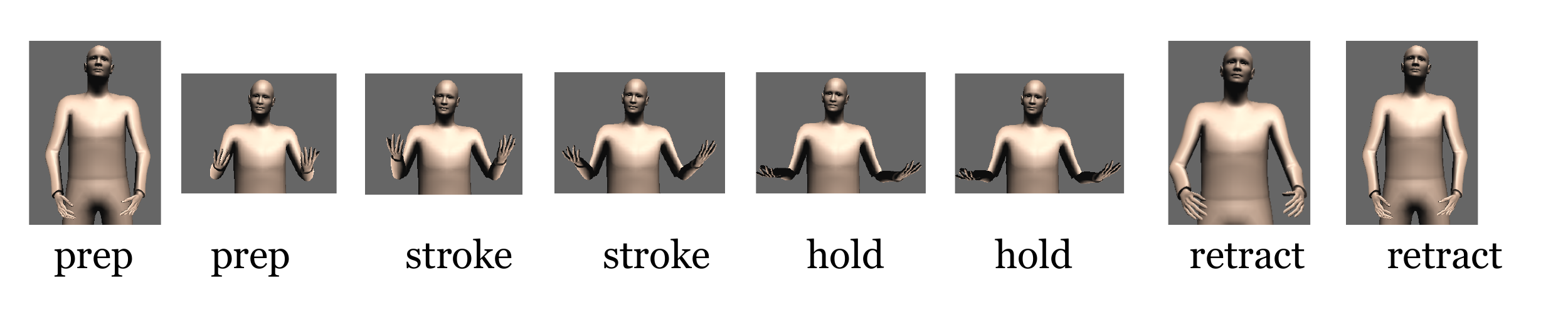}
\vspace{-0.2in}
\caption{Prep, stroke, hold and retract phases of gesture Cup\_Horizontal. \label{fig:gest-phase}}
\end{center}
\vspace{-0.3in}
\end{figure}

\begin{wrapfigure}{r}{2.6in}
\vspace{-0.2in}
\begin{center}
\begin{scriptsize}  
\begin{tabular}{|rp{2.4in}|}
\hline
A1: &[1.90s](Cup, RH 0.46s) Hey, do you remember [3.17s](PointingAbstract, RH 0.37s) that day? It was a [4.97s](Cup\_Horizontal, 2H 0.57s) work day, I remember there was some big event [7.23s](SweepSide1, RH 0.35s) going on.\\
B1: &Yeah, that day was the start of [9.43s](Cup\_Down\_alt, 2H 0.21s) the G20 summit. It's an event that happens [12.55s](CupBeats\_Small, 2H 0.37s) every year.\\
A2: &Oh yeah, [14.2s](Cup\_Vert, RH 0.54s) right, it's that meeting where 20 of the leaders of the world [17.31s](\textbf{Regressive, RH 1.14s}) come together. They talk about how to run their governments [20.72s](\textbf{Cup, RH 0.46s}) effectively.\\
B2: &Yeah, [22.08s](\textbf{Cup\_Up, 2H 0.34s}) exactly. There were many leaders [24.38s](\textbf{Regressive, LH 1.14s}/Eruptive, LH 0.76s) coming together. They had some pretty [26.77s](\textbf{WeighOptions, 2H 0.6s}) different ideas about what's the best way to [29.13s]*(\textbf{Cup, RH 0.46s}/ShortProgressive, RH 0.38s) run a government. \\
A3: &And [30.25s]*(PointingAbstract, RH 0.37s) the people who follow the governments also have [32.56s](\textbf{WeighOptions, 2H 0.6s}/Cup, 2H 0.46s) different ideas. Whenever [34.67s](\textbf{Cup\_Up, 2H 0.34s}/Dismiss, 2H 0.47s) world leaders meet, there will be protesters expressing [37.80s](Away, 2H 0.4s) different opinions. I remember the [39.87s]*(Reject, RH 0.44s) protest that happened just [41.28s](SideArc, 2H 0.57s) along the street where we work.\\
B3: &......
\\
\hline
\end{tabular}
\caption{Sample blog story dialog with gesture annotations (two versions). Format of annotation: [gesture stroke begin time](gesture name, hand use, gesture stroke duration). \label{sample_blog_dialog_gesture}}
\end{scriptsize}  
\end{center}
\vspace{-0.4in}
\end{wrapfigure}

We construct the dialogs for the two 
IVAs manually from the monolog webblogs. We then generate audio for the 
utterances of each IVA
using the AT\&T Text to Speech engine (female voice Crystal and male voice Mike).  
We annotate the dialogs with a general, underspecified gestural
representation, specifying both potential gestures and gesture
placements.  This representation allows us to procedurally generate
hundreds of possible combinations of story co-tellings varying both
gestural performance (personality) and adaptation.  Annotators can
insert a gesture when the dialog introduces new concepts, and add
gesture adaptation (mimicry) when there are repetitions or
confirmations in the dialog.  The decisions of where to insert a
gesture and which gesture to insert are mainly subjective. We use
gestures from a database of 271 motion captured gestures, including
metaphoric, iconic, deictic and beat gestures. Fig.~\ref{fig:gest-phase}
illustrates how every gesture can be
generated to include up to 4 phases~\cite{kita98}:

\begin{itemize}
\item prep: move arms from default resting position or the end point of the last gesture to the start position of the stroke
\item stroke: perform the movement  that conveys most of the gesture's meaning
\item hold: remain at the final position in the stroke
\item retract: move arms from the previous position to a default resting position
\end{itemize}

Fig.~\ref{sample_blog_dialog_gesture} shows the first 5 turns of the
protest story annotated with gestures. The timing information of the
gestures comes from the TTS audio timeline. Each gesture
annotation contains information in the following format: [gesture
  stroke begin time](gesture name, hand use, stroke duration). For
example, in the first gesture ``[1.90s](Cup, RH 0.46s)'', gesture
stroke begins at 1.9 seconds of the dialog audio, it is a Cup gesture,
uses the right hand, and the gesture stroke lasts 0.46
seconds. Research has shown that people prefer gestures occurring
earlier than the accompanying speech ~\cite{wang2013influence}. Thus
in this annotation, a gesture stroke is positioned 0.2 seconds before
the beginning of the gesture's following word. For example, the first
word after gesture ``Cup'' is ``Hey'', it begins at 2.1 seconds, then
the stroke of gesture ``Cup'' begins at 1.9 seconds.

Our gesture annotation does not specify features associated with
particular gestures (i.e. gesture amplitude, direction and speed). But
these features can be easily adjusted in our animation software,
which can vary the amplitude,
direction and speed. The default gesture annotation frequency is
designed for extraverts, with a gesture rate of 1 - 2 gestures per
sentence. For an introverted agent, a lower gesture rate is
achieved by removing some of the gestures.  In this way, both
speakers' gestural performance can vary from introverted to
extraverted using the whole scale of parameter values for every
parameter.


In addition, we can also vary gestural adaptation in the annotation.
In extravert \& extravert gestural adaptation (based on the model
described in~\cite{tolinsgestural1}), two extraverts move together
towards a more extraverted personality.  Gesture rate is increased by
adding extra gestures (marked with an asterisk ``*''). Specific
gestures are copied as part of adaptation, especially when the
co-telling involves repetition and confirmation. Gestures in bold
indicate copying of gesture form (adaptation), gestures after the
slash ``/'' are non-adapted. Combined with personality variations for
gestures described in the previous paragraph, it is possible to
produce combinations of two agents with any level of extraversion
engaged in a conversation with or without gestural adaptation.

\subsubsection{Stimulus Construction}
\label{sec:stimulus-construction}

\begin{wrapfigure}{r}{2.5in}
\vspace{-0.6in}
\begin{center}
\includegraphics[width=2.5in]{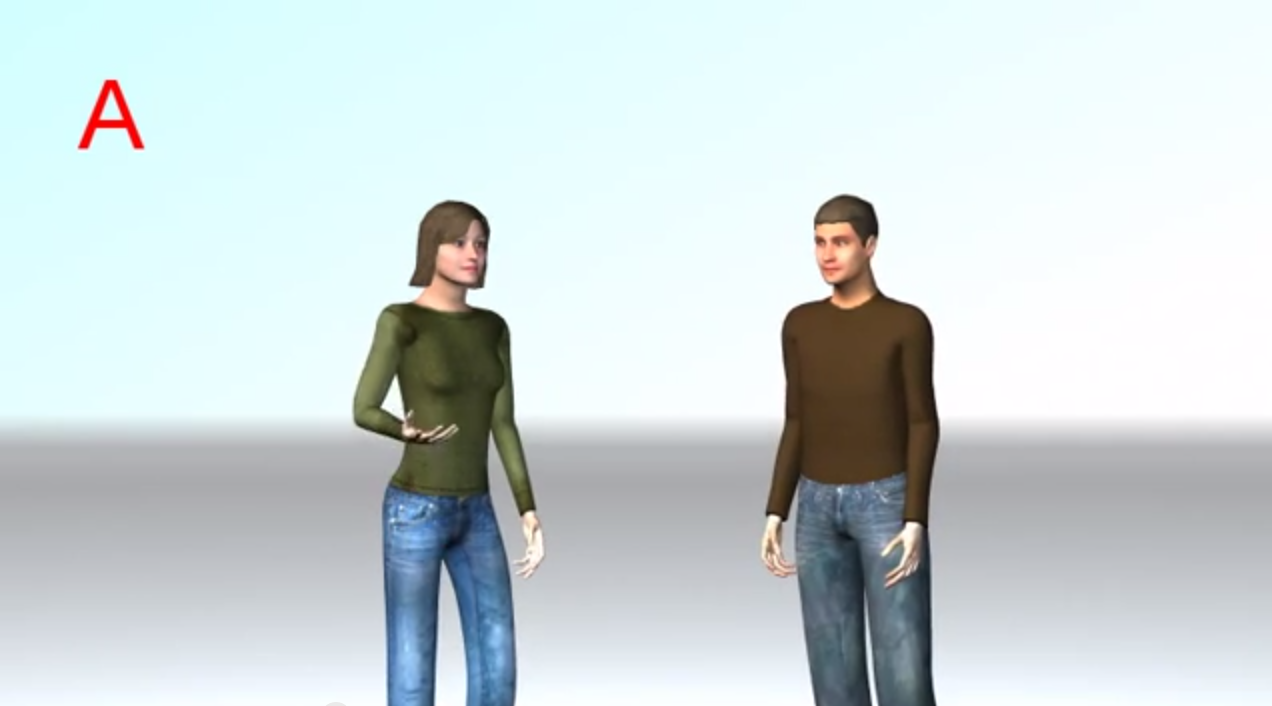}
\caption{A snapshot of the experimental stimuli.\label{stimuli-pic}}
\end{center}
\vspace{-0.4in}
\end{wrapfigure}


We currently have 50 annotated story dialogs. In this experiment, we 
use four stories with different subject matter: protest,
pet, storm and gardening, as illustrated in
Fig.~\ref{protest_blog_dialog}
and Fig.~\ref{pet_blog_dialog}.
Fig. \ref{stimuli-pic} shows a screenshot of
the stimuli.  We use our own animation software to generate the
stimuli based on the specified gesture script. This software uses
motion captured data for the wrist path, hand shape and hand
orientation for each gesture stroke, motion captured data for body
movement, and spline based interpolation for preparation and
retractions.  It also uses simplified physical simulation to add
nuance to the motion.  A gesture contains up to 4
phases: prep, stroke, hold and retract: we insert a hold and connecting prep between two strokes if they are less than 2.5 seconds away from each other. Otherwise, we insert a retraction.

The animation software takes as input scripts specifying gesture sequences, along with modifying edits (specifying features such as gesture amplitude, direction and speed), and produces an animation meeting the constraints as output. This is exported as a bvh file, that is then imported into Maya for rendering on the final model. In the video, two IVAs stand almost face-to-face, but each has an $55^{\circ}$ angle ``cheat"  towards the audience, as is commonly used in stage performances.
We also add background body movements~\cite{luo2009augmenting} and head
rotation movements for both agents.  Both of these are kept constant
for each stimuli pair.  



\section{Experiment Method}
\label{exp-sec}

\begin{wrapfigure}{r}{2.5in}
\vspace{-1in}
\begin{center}
\begin{scriptsize}  
\begin{tabular}{|rp{2.3in}|}
\hline
\multicolumn{2}{|c|}{\bf Pet Story} \\
\hline
A1: & I have always felt like I was a dog person but our two cats are great.  They are much more low maintenance than dogs are. \\

B1: & Yeah, I'm really glad we got our first one at a no-kill shelter. \\

A2: & I had wanted a little kitty, but the only baby kitten they had scratched the crap out of me the minute I picked it up so that was a big ``NO''. \\

B2: & Well, the no-kill shelter also had what they called ``teenagers'', which were cats around four to six months old...a bit bigger than the little kitties. \\

A3: & Oh yeah, I saw those ``teenagers''.  They weren't exactly adults, but they were a bit bigger than the little kittens. \\

B3: & Yeah one of them really stood out to me then-- mostly because she jumped up on a shelf behind us and smacked me in the head with her paw. \\

A4: & Yeah, we definitely had a winner! \\

B4: & I had no idea how much personality a cat can have. Our first kitty loves playing. She will play until she is out of breath. \\

A5: & Yeah, and then after playing for a long time she likes to look at you like she's saying, ``Just give me a minute, I'll get my breath back and be good to go.''  \\

B5: & Sometimes I wish I had that much enthusiasm for anything in my life. \\

A6: & Yeah, me too.  Man, she has so much enthusiasm for chasing string too! To her it's the best thing ever. Well ok, maybe it runs a close second to hair scrunchies.  \\

B6: & Oh I love playing fetch with her with hair scrunchies! \\

A7: & Yeah, you can just throw the scrunchies down the stairs and she runs at top speed to fetch them. And she always does this until she's out of breath! \\

B7: & If only I could work out that hard before I was out of breath... I'd probably be thinner. \\
\hline 
\end{tabular}
\caption{Pet Dialogue, with fixed level of linguistic
  adaptation. \label{pet_blog_dialog}}
\end{scriptsize}  
\end{center}
\vspace{-0.5in}
\end{wrapfigure}

We conduct two separate experiments, one on personality variation
during co-telling a story, and the second using the same
personalities but with and without adaptation.
\subsubsection{Experiment 1: Personality Variation.}
We prepared two versions of the video of the story co-telling for each of the four stories, one where the
female is extraverted (higher values for gesture rate, gesture expanse,
  height, outwardness, speed and scale) and the male is introverted (lower values for those gesture features) and one where only the genders (virtual agent model and voice) of the agents are switched. The dialogue scripts and corresponding gesture forms
do not vary from one co-telling to
another. This results in 8 video stimuli for four stories. 

We conducted a between-subjects experiment on Mechanical Turk
where we first ask Turkers to answer the TIPI~\cite{Goslingetal03} personality survey for
themselves, and then answer it for only one of the agents in the video,
after watching the video as many times as they like. Thus for each video stimulus, there are two surveys. 
We ran our 16 surveys as 16 HITs (Human Intelligence Tasks) on Mechanical Turk, requesting 20 subjects per HIT (each worker can only do one of the tasks), which results in 320 judgements. The average completion time for the
8 HITs on Mechanical Turk was 5 minutes 15 seconds. The average
stimulus length was 1 minute 32 seconds. Since the survey is hosted outside Mechanical Turk, sometimes we get more than 20 subjects for each HIT.  

\subsubsection{Experiment 2: Gestural Adaptation.}
For the adaptive experiment, both agents are designed to be extroverted. We chose to use two extraverted agents because we have foundations from previous work showing the adaptation model between two extraverted speakers~\cite{tolinsgestural1} (where both agents become more extraverted). 
We use only a part of each
story for one experimental task. The stimuli for one task has two
variations: adapted and non-adapted. Both stimuli use the same audio,
contain 2 to 4 dialog turns with the same gestures as an introduction
to the story (which we refer to as context), and the next (and last) dialog turn
with gesture adaptation or without gesture adaptation (which we refer
to as response). Adaptation only begins to occur in the last dialog turn. 
In this way, subjects can get to know the story through the context, and
compare the responses to decide whether they like the adapted or non-adapted version. 

\begin{wrapfigure}{r}{3in}
\vspace{-0.5in}
\begin{center}
\includegraphics[width=3in]{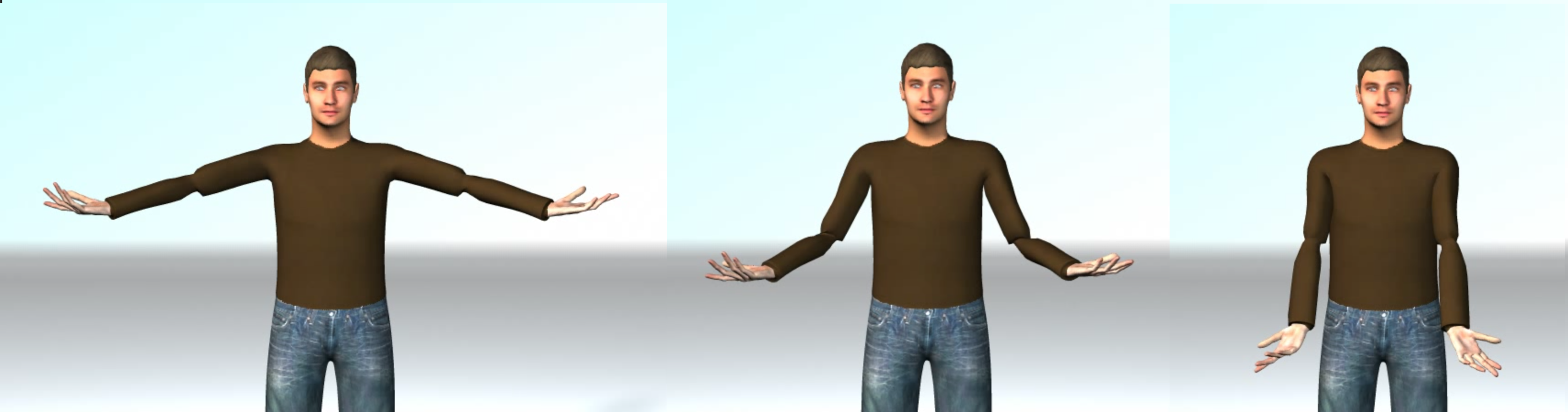}
\caption{Virtual agent with different gesture expanse and height for the same gesture.\label{gest-fig}}
\end{center}
\vspace{-0.5in}
\end{wrapfigure} 

\begin{itemize}
\item Non-adapted: In the last dialog turn, the extraverted agent maintains
  his or her gesture rate (1 - 2 gestures per sentence), expanse,
  height, outwardness, speed and scale. There is no copying of
  specific gestures.
 
\item Adapted: In the last dialog turn, the extraverted agent increases the
  gesture rate (1 - 3 gestures per sentence), expanse (18 cm
  further from center), height (10 cm higher), outwardness (10 cm more outward),
  speed (1.25 times faster) and scale (1.5 times
  larger). Fig.~\ref{gest-fig} shows the same gesture with different
  expanses and heights. In the adapted version, specific gestures are
  copied (e.g. gestures in bold font in
  Fig.\ref{sample_blog_dialog_gesture}).
\end{itemize}

Thus every story has two versions. One version ends with the female
agent's response, another ends with the male agent's response. 
For example, Garden
ABA has three turns, ending with the female agent adapting to the
male, and Garden ABAB has four turns, ending with the male agent
adapting to the female. Every version consists of two conditions
(adapted and non-adapted versions) and a short survey. The order of
the two conditions is random for every participant. But there is a
letter mark assigned to every video for easy reference (see
Fig.~\ref{stimuli-pic}).

Subjects are asked to watch the two stimuli first, and then finish the
survey. Subjects are told that the audio of the two videos is the
same, but only the last few gestures of the female/male agent are
different. Subjects are also advised to watch the video as many times
as they want.  The survey has two questions: (1) Which video is a
better story co-telling based on the gestures? (2) Please explain the
reason behind your choice to the previous question (which we refer to
as the ``why'' question). 
Our primary aim is to determine whether people perceive the adaptation
and whether it makes a better story.

We ran our 8 tasks for 4 stories as 8 HITs on Mechanical Turk,
requesting 25 subjects per task.  The average completion time for the
8 tasks on Mechanical Turk was 2 minutes 53 seconds. The average
stimulus length was 35.3 seconds. This means that, on average, a
subject spent 1 minute 43 seconds answering the questions. We removed subjects 
who failed to state their reasons of preference in the "why" question.

\section{Experimental Results}
\label{res-sec}

\subsection{Personality Results}
\label{pers-results}

\begin{wraptable}{r}{1.6in}
\vspace{-0.6in}
    \begin{tabular}{p{0.5in}|p{.1in}p{0.4in}p{0.4in}}
    \toprule 
Story & &Intro-Agent & Extra-Agent \\
        \midrule
    Garden && 4.2 & 5.4 \\
    Pet   & & 4.7 & 5.0 \\
    Protest && 4.2 & 5.3 \\
    Storm && 3.7 & 5.7 \\
    \bottomrule
    \end{tabular}%
  \caption{Experiment results: participant evaluated extraversion scores  (range from 1 - 7, with 1 being the most introverted and 7 being the most extraverted).   \label{tab:personality-results}}
\vspace{-0.4in}
\end{wraptable}
We conducted a three-way ANOVA with agent intended personality, agent
gender and story as independent variables and perceived agent
personality as the dependent variable. 
See Table~\ref{tab:personality-results}. The results show 
that subjects clearly perceive the intended extraverted or
intended introverted personality of the
two agents (F = 67.1, p $<$ .001). There is no main effect
for story (as intended in our design), but there is an interaction
effect between story and intended personality, with the introverted
agent in the storm story being seen as much more introverted than
in the other stories (F= 7.5, p $<$ .001). There is no significant
variation by agent gender (F = 2.3, p = .14).

Since previous work suggests that personality is perceived for an
agent along all Big Five dimensions whether it is designed to be
manifest or not \cite{MairesseWalker11,Liuetal13}, we also conducted a
two-way ANOVA by story and agent intended personality for the other 4
traits. There are no significant differences for Conscientiousness, or
Openness. However Introverted agents are seen as more agreeable (p =
.008) and more emotionally stable (p = .016). There were no significant
differences by story except that both agents in the Storm story were
seen as less open, presumably because the content of the story is
about how scary the storm is.



  

\subsection{Adaption Results}
\begin{wraptable}{r}{2.4in}
\vspace{-0.9in}
  \centering
    \begin{tabular}{p{1in}p{0.3in}p{0.3in}p{0.3in}p{0.3in}}
    \toprule
    Story Version & \#A & \#NA & \%A & \%NA \\
    \midrule
    Garden ABA & 11    & 9     & 55\%  & 45\% \\
    Garden ABAB & 20    & 2     & 91\%  & 9\% \\
    Pet ABABA & 10    & 13    & 43\%  & 57\% \\
    Pet ABABAB & 19    & 5     & 79\%  & 21\% \\
    Protest ABAB & 8     & 11    & 42\%  & 58\% \\
    Protest ABABA & 11    & 11    & 50\%  & 50\% \\
    Storm ABABA & 16    & 4     & 80\%  & 20\% \\
    Storm ABABAB & 14    & 5     & 74\%  & 26\% \\
    \midrule
    Total & 109   & 60    & 64\%  & 36\% \\
    \bottomrule
    \end{tabular}%
    \caption{Experiment results: number and percentage of subjects who preferred the adapted (A) stimulus and the non-adapted (NA) stimulus. The letters in the story version refer to dialog turns by speaker A or B. For example, ABA means A takes dialog turns 1 and 3 in the stimuli, while B takes dialog turn 2.}
     \label{tab:results}%
\vspace{-0.3in}
\end{wraptable}

The results in Table~\ref{tab:results} show that across all the 
videos, the mean percentage of people who preferred the adapted
version was $64\%$ ($19\%$ standard deviation), which is marginally
better than a predicted preference of $50\%$, $t(7) = 2.15, p = .07$.
Analysis of participants' descriptions of why they preferred one video
over another shows 4 distinct categories of reasons of why people made
their choices (see Table~\ref{tab:results-why}).

Subjects who preferred the adapted versions said
that the gestures fit the dialog better 
(``adapted good gestures'' in Table~\ref{tab:results-why}): the
subjects stated that the adapted versions had gestures that ``flowed
better with the words'', were ``more natural'', ``more appropriate to
what he said'', and ``relevant to the dialog'', and that they ``could
imagine a friend making various hand gestures similar'' to the ones in
the story. Another reason was that gestures were ``more animated''
(``adapted animated''): the
adapted version had ``more hand gestures'', and the
agent ``used his arms more'', ``gestured more'', and ``was much more
alive''. In contrast, in the non-adapted version, the agent ``seemed
very bored'' and ``wanted to end the conversation''. This
indicates that the subjects preferred agents with a
higher gesture rate. Ten subjects
commented on the expanse, height, scale and speed of the gestures: they
chose the adapted version because the agent ``gestured higher in the
air'', ``making wider, grander gestures'' that were ``more expansive''
and ``bigger''. And in the non-adapted version, the gestures were
``too slow''. However, there was no comment about the copying of
gestures, possibly because copying
was less obvious when the expanse and height of the
gestures changed in the adapted version.

Among those who preferred the non-adapted versions of the stories,
one reason was that the gestures fit the dialog better
(``non-adapted good gestures'' in
Table~\ref{tab:results-why}): the subjects stated that the gestures in
the non-adapted version ``went a lot better with what she was saying''
and were ``more appropriate''. Another reason is that the gestures were
``more realistic'' (``non-adapted realistic'') :
subjects didn't like the gestures being ``too animated'', or ``too busy'',
nor did they like the agents ``showing way too much emotions'' or ``looking like
she is exercising''. That is, too much animation can be seen as unrealistic.


\begin{wraptable}{r}{2.6in}
\vspace{-0.3in}
  \centering
    \begin{tabular}{p{1in}p{0.3in}p{0.3in}p{0.4in}p{0.4in}}
    \toprule
Story Version& \%A good gest  & \%NA good gest   & \%A animated & \%NA realistic \\
    \midrule
    Garden ABA & 30\%  & 30\%  & 20\%  & 30\% \\
    Garden ABAB & 41\%  & 9\%   & 59\%  & 0\% \\
    Pet ABABA & 22\%  & 43\%  & 13\%  & 9\% \\
    Pet ABABAB & 54\%  & 13\%  & 33\%  & 0\% \\
    Protest ABAB & 21\%  & 32\%  & 26\%  & 0\% \\
    Protest ABABA & 27\%  & 32\%  & 23\%  & 9\% \\
    Storm ABABA & 20\%  & 15\%  & 45\%  & 0\% \\
    Storm ABABAB & 32\%  & 21\%  & 47\%  & 0\% \\
     \midrule
    Total & 31\%  & 24\%  & 33\%  & 6\% \\
    \bottomrule
    \end{tabular}%
    \caption{Answers to the second survey question (``why'' question) classified into categories. Note that one subject could belong to none or multiple categories, so the percentages for each line don't add up tp 100\%.}
      \label{tab:results-why}%

\vspace{-0.6in}
\end{wraptable}

The percentages of the subjects that had comments related to those 4
categories are in Table~\ref{tab:results-why}. In 7 out of 8
tasks, there were more subjects who preferred the adapted version because
it was animated at the right level (e.g. animated enough, but not too animated). If we only
consider the ``animated'' factor in deciding which is a better
stimulus, 84\% of the subjects preferred the adapted version. 

%

\section{Discussion and Future Work}
\label{conc-sec}

To our knowledge this is the first time that it has been shown that
subjects perceive differences in agent personality during a storytelling task, and that adaptive gestural behavior during storytelling
is positively perceived.  We re-use natural personal narratives that
are rendered dialogically, so that two IVAs co-tell the story.

It is obvious that being able to adapt is a key part of being more
human-like. There are attempts to integrate language adaptation within
natural language generation \cite{spud} and research has shown that
human bystanders perceive linguistic adaptation positively
\cite{entrainment2014hu}. However, this is the first experiment to
demonstrate a positive effect for gestural adaptation.

Recent work on gesture generation has focused largely on iconic
gesture generation. For example, Bergmann and Kopp
\cite{bergmann2009increasing} present a model that allows virtual
agents to automatically select the content and derive the form of
coordinated language and iconic gestures. Luo et
al.~\cite{luo2009augmenting} also presents an effective algorithm for
adding full body postural movement to animation sequences of arm
gestures.
More generally, current systems generally select gestures using either a
text-to-gesture or concept-to-gesture mapping. Text-to-gesture systems, such as
VHP~\cite{vhp-noma2000design}, may have a limited number of gestures (only 7
in this case) and limited gesture placement options, but the alignment
of speech content and gestures are more accurate. Concept-to-gesture
systems such as PPP~\cite{ppp-andre1997wip},
AC and BEAT~\cite{cassell2004beat} 
defines general rules for gesture insertion based on linguistic
components. For example, iconic gestures are triggered by words with spatial or concrete context (e.g. ``check''). These kind of
systems have more gestures, but the gesture placement largely depends
on general rules derived from literature, thus the accuracy is not
guaranteed.  An alternative approach learns a personalized statistical model that predicts a gesture given the text to be spoken and a model that captures an individual's gesturing preferences~\cite{neff2008gesture}.
None of these models adequately address the production of gesture for dialogues, where a process of co-adaptation will modulate both the type of gesture chosen and the specific form of that gesture (e.g. its size).  This current work aims to provide a basis for developing such models.

Gratch investigates creating rapport with virtual agents using gesture
adaptation mainly focused on head gestures and posture shifts (while
ours focused on hand gestures), and used real human movements as
control ~\cite{gratch2007creating}.  Our adaptation stimuli are more
similar to Endrass et al.~\cite{DBLP:journals/aamas/EndrassARN13}. To
investigate culture-related aspects of behavior for virtual
characters, they chose prototypical body postures from corpora for
German and Japanese cultural background, embodied those postures in a
two-agent dialogs, and asked subjects from German and Japanese
cultural background to evaluate the dialogs.

In future work, we aim to test the expression of personality
and adaptivity with different personality combinations. Our ultimate
goal is to automatically convert monologic blog stories to dialogs with
both linguistic and gestural adaptation. Experimental exploration,
such as undertaken here, is crucial for formulating models of gesture
generation that correctly incorporate personality and adaptation.


\begin{thebibliography}{4}

	\bibitem{nisbett1980trait} R.E. Nisbett.
\newblock {The trait construct in lay and professional psychology}.
\newblock {\em Retrospections on social psychology}, pages 109--130, 1980.

	\bibitem{Mehletal06} M.~R. Mehl, S.~D. Gosling, and J.~W. Pennebaker.
\newblock Personality in its natural habitat: Manifestations and implicit folk
  theories of personality in daily life.
\newblock {\em Journal of Personality and Social Psychology}, 90:862--877,
  2006.

	\bibitem{Norman63} W.~T. Norman.
\newblock Toward an adequate taxonomy of personality attributes: Replicated
  factor structure in peer nomination personality rating.
\newblock {\em Journal of Abnormal and Social Psychology}, 66:574--583, 1963.

	\bibitem{BickmoreSchulman06} T.~Bickmore and D.~Schulman.
\newblock {The comforting presence of relational agents}.
\newblock In {\em CHI'06 extended abstracts on Human factors in computing
  systems}, pages 550--555. ACM, 2006.

%

	\bibitem{Hartmann2005} B.~Hartmann, M.~Mancini, and C.~Pelachaud.
\newblock {Implementing expressive gesture synthesis for embodied
  conversational agents}.
\newblock In {\em {Proc.~Gesture Workshop 2005}}, volume 3881 of {\em LNAI},
  pages 45--55, Berlin; Heidelberg, 2006.

	\bibitem{kopp04} S.~Kopp and I.~Wachsmuth.
\newblock {Synthesizing multimodal utterances for conversational agents}.
\newblock {\em Computer Animation and Virtual Worlds}, 15:39--52, 2004.

	\bibitem{thie08} M.~Thiebaux, A.~Marshall, S.~Marsella, and M.~Kallman.
\newblock Smartbody: Behavior realization for embodied conversational agents.
\newblock In {\em Proc. of 7th Int. Conf. on Autonomous Agents and Multiagent
  Systems (AAMAS 2008)}, pages 151--158, 2008.

%

	\bibitem{helo09} A.~Heloir and M.~Kipp.
\newblock {EMBR--A Realtime Animation Engine for Interactive Embodied Agents}.
\newblock In {\em Intelligent Virtual Agents 09}, pages 393--404. Springer,
  2009.

	\bibitem{Andreetal99} E.~Andr\'e, M.~Klesen, P.~Gebhard, S.~Allen, and T.~Rist.
\newblock Integrating models of personality and emotions into lifelike
  characters.
\newblock In {\em Proc.~of the Workshop on Affect in Interactions -
  Towards a new Generation of Interfaces}, pages 136--149, 1999.

	\bibitem{Ruttkayetal04} Z. Ruttkay, C. Dormann, and H. Noot.
\newblock Embodied conversational agents on a common ground.
\newblock {\em From
  brows to trust: evaluating embodied conversational agents}, chapter~2, pages
  27--66. Kluwer Academic Publishers, Norwell, MA, 2004.

%

	\bibitem{MoonNass96} Y.~Moon and C.~Nass.
\newblock How ``real'' are computer personalities?: Psychological responses to
  personality types in human-computer interaction.
\newblock {\em Communication Research}, 1996.

	\bibitem{DBLP:journals/aamas/EndrassARN13} B. Endra{\ss}, E. Andr{\'{e}}, M. Rehm, and Y.I. Nakano.
\newblock Investigating culture-related aspects of behavior for virtual
  characters.
\newblock {\em Autonomous Agents and Multi-Agent Systems}, 27(2):277--304,
  2013.

	\bibitem{vhp-noma2000design} T. Noma, N.~I. Badler, and L. Zhao.
\newblock Design of a virtual human presenter.
\newblock {\em Center for Human Modeling and Simulation}, page~75, 2000.

	\bibitem{tapus2008a} A.~Tapus, C.~Tapus, and M.J. Mataric.
\newblock {User robot personality matching and assistive robot behavior
  adaptation for post-stroke rehabilitation therapy}.
\newblock {\em Intelligent Service Robotics}, 1(2):169--183, 2008.

	\bibitem{Wangetal05} N. Wang, W.~L. Johnson, R.~E. Mayer, P. Rizzo, E. Shaw, and
  H. Collins.
\newblock The politeness effect: Pedagogical agents and learning gains.
\newblock {\em Frontiers in AI and Applications}, 2005.

	\bibitem{MairesseWalker11} F. Mairesse and M.~A. Walker.
\newblock Controlling user perceptions of linguistic style: Trainable
  generation of personality traits.
\newblock {\em Computational Linguistics}, 2011.

	\bibitem{Neffetal10} M.~Neff, Y.~Wang, R.~Abbott, and M.~Walker.
\newblock Evaluating the effect of gesture and language on personality
  perception in conversational agents.
\newblock In {\em IVA}, pages 222--235. Springer, 2010.

%

	\bibitem{Beeetal10} N.~Bee, C.~Pollock, E.~Andr{\'e}, and M.~Walker.
\newblock Bossy or wimpy: expressing social dominance by combining gaze and
  linguistic behaviors.
\newblock In {\em Intelligent Virtual Agents}, 2010.

	\bibitem{Neffetal11} M.~Neff, N.~Toothman, R.~Bowmani, J.E. Fox~Tree, and Walker~M. A.
\newblock Dont scratch! self-adaptors reflect emotional stability.
\newblock In {\em IVA}, volume 6895. Springer, 2011.

	\bibitem{Foxtree99} J.~E. Fox~Tree.
\newblock Listening in on monologues and dialogues.
\newblock {\em Discourse Processes}, 27:35--53, 1999.

%

	\bibitem{ParrillKimbara06} F.~Parrill and I.~Kimbara.
\newblock Seeing and hearing double: The influence of mimicry in speech and
  gesture on observers.
\newblock {\em Journal of Nonverbal Behavior}, 30:157--166, 2006.

%

	\bibitem{BailensonNick05} J.N.~Bailenson and N.~Yee.
\newblock Digital chameleons: Automatic assimilation of nonverbal gestures in
  immersive virtual environments.
\newblock {\em Psychological Science}, 16(10):814--819, 2005.

	\bibitem{tolins2014addressee} J. Tolins and J. E.~Fox Tree.
\newblock Addressee backchannels steer narrative development.
\newblock {\em Journal of Pragmatics}, 70:152--164, 2014.

	\bibitem{thorne2007channeling} A. Thorne, N. Korobov, and E.~M. Morgan.
\newblock Channeling identity: A study of storytelling in conversations between
  introverted and extraverted friends.
\newblock {\em Journal of research in personality}, 41(5):1008--1031, 2007.

%

	\bibitem{GordonSwanson09} A. Gordon and R. Swanson.
\newblock Identifying personal stories in millions of weblog entries.
\newblock In {\em 3rd Int.~Conference on Weblogs and Social Media,
  Data Challenge Workshop, San Jose, CA}, 2009.

	\bibitem{kita98} S.~KIita, I.~{Van Gijn}, and H.~{Van Der Hulst}.
\newblock Movement phase in signs and co-speech gestures, and their
  transcriptions by human coders.
\newblock In {\em Proc.~of the Int.~Gesture Workshop on Gesture
  and Sign Language in Human-Computer Interaction},
  Springer-Verlag, 1998.

	\bibitem{wang2013influence} Y. Wang and M. Neff.
\newblock The influence of prosody on the requirements for gesture-text
  alignment.
\newblock In {\em IVA}, pages 180--188. Springer, 2013.

	\bibitem{tolinsgestural1} J. Tolins, K. Liu, Y. Wang, J. E.~Fox Tree, M. Walker, and
  M. Neff.
\newblock Gestural adaptation in extravert-introvert pairs and implications for
  ivas.
\newblock In {\em IVA}, page 484. Springer.

	\bibitem{luo2009augmenting} P. Luo, M. Kipp, and M. Neff.
\newblock Augmenting gesture animation with motion capture data to provide
  full-body engagement.
\newblock In {\em IVA}, pages 405--417. Springer, 2009.

	\bibitem{Goslingetal03} S.~D. Gosling, P.~J. Rentfrow, and W.~B. Swann.
\newblock A very brief measure of the big five personality domains.
\newblock {\em Journal of Research in Personality}, 37:504--528, 2003.

	\bibitem{Liuetal13} K. Liu, J. Tolins, J. E.~Fox Tree, M. Walker, and M. Neff.
\newblock Judging iva personality using an open-ended question.
\newblock In {\em IVA}, pages 396--405. Springer, 2013.

	\bibitem{spud} H. Buschmeier, K. Bergmann, and S. Kopp.
\newblock An alignment-capable microplanner for natural language generation.
\newblock In {\em Proc.~of the 12th European Workshop on Natural Language
  Generation}, pages 82--89. ACL, 2009.

%

	\bibitem{entrainment2014hu} Z. Hu, G. Halberg, C. Jimenez, and M. Walker.
\newblock Entrainment in pedestrian direction giving: How many kinds of
  entrainment?
\newblock In {\em IWSDS}, 2014.

%

	\bibitem{bergmann2009increasing} K. Bergmann and S. Kopp.
\newblock Increasing the expressiveness of virtual agents: autonomous
  generation of speech and gesture for spatial description tasks.
\newblock In {\em Proc.~of The 8th Int.~Conference on Autonomous
  Agents and Multiagent Systems-Volume 1}, pages 361--368. 2009.

	\bibitem{ppp-andre1997wip} E.~Andr{\'e}, J.~M{\"u}ller, and T.~Rist.
\newblock Wip/ppp: Automatic generation of personalized multimedia
  presentations.
\newblock In {\em Proc.~of the fourth ACM international conference on
  Multimedia}, pages 407--408. ACM, 1997.

%

	\bibitem{cassell2004beat} J. Cassell, H. H. Vilhj{\'a}lmsson, and T. Bickmore.
\newblock Beat: the behavior expression animation toolkit.
\newblock In {\em Life-Like Characters}, pages 163--185. Springer, 2004.

	\bibitem{neff2008gesture} M. Neff, M. Kipp, I. Albrecht, H. P. Seidel
\newblock Gesture modeling and animation based on a probabilistic re-creation of speaker style.
\newblock In {\em ACM Transactions on Graphics (TOG)}. ACM, 2008.

	\bibitem{gratch2007creating} J. Gratch, N. Wang, J. Gerten, E. Fast, and R. Duffy.
\newblock Creating rapport with virtual agents.
\newblock In {\em IVA}, pages 125--138. Springer, 2007.

\end{thebibliography}
\end{document}